# Microscopic mechanism of tunable thermal conductivity in carbon nanotube-geopolymer nanocomposites

*Wenkai Liu [a], Ling Qin [b], C. Y. Zhao [a], Shenghong Ju [a,]\**

a China-UK Low Carbon College, Shanghai Jiao Tong University, Shanghai, 201306, China.

b SPIC Guizhou Jinyuan Co., Ltd., Guiyang, Guizhou, 550081, China.

ABSTRACT. Geopolymer has been considered as a green and low-carbon material with great potential application due to its simple synthesis process, environmental protection, excellent mechanical properties, good chemical resistance and durability. In this work, the molecular dynamics simulation is employed to investigate the effect of the size, content and distribution of carbon nanotubes on the thermal conductivity of geopolymer nanocomposites, and the microscopic mechanism is analyzed by the phonon density of states, phonon participation ratio and spectral thermal conductivity, etc. The results show that there is a significant size effect in geopolymer nanocomposites system due to the carbon nanotubes. In addition, when the content of carbon nanotubes is 16.5%, the thermal conductivity in carbon nanotubes vertical axial direction (4.85

W/(m·k)) increases 125.6% compared with the system without carbon nanotubes (2.15 W/(m·k)). However, the thermal conductivity in carbon nanotubes vertical axial direction (1.25 W/(m·k)) decreases 41.9%, which is mainly due to the interfacial thermal resistance and phonon scattering at the interfaces. The above results provide theoretical guidance for the tunable thermal conductivity in carbon nanotube-geopolymer nanocomposites.

1. INTRODUCTION

The geopolymer (GP)[1] composed of three-dimensional (3D) amorphous networks with $[AlO_4]^{5-}$ and $[SiO_4]^{4-}$ tetrahedrons[2] has shown excellent mechanical properties, good chemical resistance, long-term durability and low shrinkage.[3] Compared with ordinary portland cement, the GP does not require a high calcination temperature, and thus the production process can reduce $CO_2$ emissions by 70-80%.[4-7] Meanwhile, the source materials for synthesizing GP are very cheap, including metakaolin, fly ash, ground granulated blast furnace slag, etc.[8] These unique properties expand the application of GP in fields of building materials,[9] aerospace materials,[10] and fire-resistance materials[11], in which the thermal conductivity of GP plays important role. Due to the nature amorphous structure, there is strong phonon scattering during the thermal transport process, leading to low thermal conductivity. However, this greatly limits the application of GP in geothermal drilling,[12] molten salt storage system,[13] nuclear waste coating[14], etc., which usually requires high thermal conductivity to improve energy efficiency, while low temperature gradients to reduce the thermal stress.[15]

At present, research progress has been reported in enhancing the thermal conductivity of GP, mainly by adding high thermal conductivity nanofillers including quartz,[16] SiC,[17] graphene nanoplatelet,[18] diamond powder,[14] carbon nanotubes (CNT),[2] etc. Subaer et al.[16] added 40 wt% quartz to the raw material and found that the thermal conductivity of GP increased by about 40%, and they attributed this higher thermal conductivity to the increased density of the composites material. Du et al.[17] investigated the effect of adding different content of silicon carbide particles (SiC) on the thermal conductivity of GP composites. It was found that 10 wt% SiC increases the thermal conductivity of GP by 19.49%, while 1 wt% SiC decreases the thermal conductivity by 7.06% compared with the unfilled GP. This was mainly because the thermal conductivity of the GP composites was determined by both filler content and porosity. Graphene nanoplatelet is also a good filler for improving the electrical and thermal conductivity of GP. Cho et al.[18] found that the thermal conductivity of the GP can be increased by 22% when adding 3 wt% of graphene nanoplatelet to potassium geopolymer. Wang et al.[14] added 15 wt% of diamond powder to GP by using 3D printing technology, which increased its thermal conductivity by 47.85%.

In addition to the reinforcing phases mentioned above, the CNT is another ideal reinforcing fillers[19], since CNT has good mechanical, electrical, chemical and thermal properties. Abbasi et al.[20] investigated the strengthening mechanism of multiwalled carbon nanotubes (MWCNT) on the compressive and flexural strength of metakaolin-based GP. It was found that adding 0.5 wt% CNT can increase the compressive strength and flexural strength of the composites by 32% and 28%, respectively. They attributed this to CNT bridging microcracks within the GP, and the number of bridged microcracks was affected by the concentration and dispersion of CNT. In addition, Saafi et al.[19] also found that MWCNT/fly ash-based geopolymer has a high piezoresistive effect, and the measured electrical impedance suddenly increases due to the generation and expansion of internal cracks. This piezoresistive mechanism provides the possibility for early detection of crack emergence in GP materials. Moreover, by coating the surface of CNT with uniform silica, Zhu et al.[2] reduced the interfacial thermal resistance and modulus mismatch between CNT and GP, while improving the dispersibility of CNT, which significantly improved the thermal conductivity of the composites.

The current researches on enhancing the thermal conductivity of geopolymer are all at experimental level, and the microscopic mechanism of the nano-enhanced phase affecting the thermal conductivity of GP composites at atomic scale is still unclear, which largely limits the further improvement of their thermal conductivity. Molecular dynamics (MD) simulations provide an effective way to explore the physical mechanism inside the geopolymer nanocomposites and essential information for multi-scale studies coupled with mechanical properties. In this work, the MD simulation is conducted to study the effect and mechanism of CNT on the thermal conductivity of carbon nanotube-geopolymer nanocomposites (CNT-GP), aiming to explore the physical tuning mechanism of the thermal conductivity of CNT-GP composites.

## 2. COMPUTATIONAL DETIALS

### 2.1. Atomic Modeling

In order to investigate the effect of system size and CNT content on the thermal conductivity of the CNT-GP nanocomposites, the atomistic structure for different size systems and CNT content systems were constructed, as shown in Table 1 and Figure 1. It is worth pointing out that we limited our study to sodium-based geopolymer and Na/Al ratio is 1 to ensure the charge balance of the system. During the atomistic generation process, the monomer structure of Poly(sialate-disiloxo) (-Si-O-Al-O-Si-O-Si-O-) and sodium atom was first generated. Then, a certain content of monomers were randomly added into a cubic box and the tolerance value of 2.0Å was set to avoid atomistic overlapping. The system was set up with periodic boundary conditions in all three directions, and was then relaxed under NVT ensemble runs for 200 ps at 300 K with time step of 1 fs. The temperature was then increased to 4000 K and subsequently decreased to 300 K at a rate of 5 K/ps to ensure the generation of amorphous network structure. The system was further equilibrated for 500 ps under NPT ensemble, followed by a NVT ensemble running for 500 ps at 300K, producing a stable GP structure. Finally, different amount of CNT was added to GP, and the CNT-GP nanocomposite system was sequentially equilibrated for 200 ps under the NPT and NVT ensembles, respectively, to obtain the stable CNT-GP nanocomposite structure. We found that the equilibrium relaxation time set in this study was sufficient for all systems to form stable structures. And in order to reduce the statistical error, three different initial structures were generated for all systems, and the standard error of the results was calculated as the error bar.

**Table 1.** Simulated CNT-GP nanocomposites systems

| Name | System size | CNT content (%) | Atoms |
| --- | --- | --- | --- |
| GP1 | 44Å×44Å×44Å | 0 | 6331 |
| GP2 | 88Å×44Å×44Å | 0 | 12662 |
| GP3 | 132Å×44Å×44Å | 0 | 18993 |
| GP4 | 176Å×44Å×44Å | 0 | 25324 |
| CNT/GP1 | 44Å×44Å×44Å | 3.3 | 6482 |
| CNT/GP2 | 88Å×44Å×44Å | 3.3 | 12968 |
| CNT/GP3 | 132Å×44Å×44Å | 3.3 | 19478 |
| CNT/GP4 | 176Å×44Å×44Å | 3.3 | 25970 |
| CNT/GP5 | 44Å×44Å×44Å | 9.9 | 6735 |
| CNT/GP6 | 88Å×44Å×44Å | 9.9 | 13606 |
| CNT/GP7 | 132Å×44Å×44Å | 9.9 | 20472 |
| CNT/GP8 | 176Å×44Å×44Å | 9.9 | 27313 |
| CNT/GP9 | 44Å×44Å×44Å | 16.5 | 7141 |
| CNT/GP10 | 88Å×44Å×44Å | 16.5 | 14207 |
| CNT/GP11 | 132Å×44Å×44Å | 16.5 | 21409 |
| CNT/GP12 | 176Å×44Å×44Å | 16.5 | 28608 |

**Figure 1.** Simulated CNT-GP nanocomposites systems. (a) Front view, (b-e) Side view. (f) Schematic diagram of the monomer structure. (g) A zoomed in image of the CNT-GP nanocomposite structure in (c).

## 2.2. Interatomic potential.

In MD simulations, the interatomic potential is crucial to accurately describe the atomistic interactions. In this work, we chose the interatomic potential that have been verified earlier for GP system with similar composition, including zeolites,[21] metakaolin[22] and geopolymer.[23] The expression of the interaction potential between M-O (M=Si, Al, Na, O) is :

$$V(r) = \frac{Cq_iq_j}{\epsilon r} + A\, exp\left(-\frac{r}{\rho}\right) - \frac{C}{r^6} + \frac{D}{r^{12}} \quad r < r_c, \tag{1}$$

where the first term is the long-range Coulomb interaction in which C is the energy conversion constant, $q_i$ and $q_j$ are the charges of atom $i$ and $j$, $\epsilon$ is the dielectric constant, $r$ is the distance between the two atoms, $r_c$ is the cutoff radius set as 12Å. The second and third terms are Buckingham interactions. Usually, the simulation system will be collapsed under the Buckingham potential at high temperature, which is mainly due to that the value of $V(r)$ is negative infinity when $r$ is infinitely small. In this work, we add the fourth repulsive term, which is widely used in other work.[24,25] The detail parameters setting of $q_i$, $q_j$, $A$, $\rho$, $C$, $D$ are shown in Table 2. In this study, only mechanical doping of CNT is considered, which is in good agreement with the actual experimental doping process, that is, there is no covalent bonding interaction between CNT and GP, but only van der Waals force interaction. So for the interaction between GP and CNT (M-C), the Lennard-Jones potential was chosen[26,27] which is expressed as,

$$V(r) = 4\varepsilon\left[\left(\frac{\sigma}{r}\right)^{12} - \left(\frac{\sigma}{r}\right)^{6}\right] \quad r < r_c, \tag{2}$$

where $\varepsilon$ is the depth of the potential well, $\sigma$ is the distance where the potential equals zero. The parameters settings of $\varepsilon$, $\sigma$ are shown in Table 2. The optimized Tersoff potential[28] was chosen to describe the C-C interactions in CNT. All simulations in this study were conducted by using LAMMPS.[29]

**Table 2.** Interatomic potential parameters

| Atoms | Charge | Atoms | Charge |
|---|---|---|---|
| Si | +2.4 | O | -1.2 |
| Al | +1.8 | C | 0 |
| Na | +0.6 | | |
| Buckingham potential parameters | | | |
| Atomic pair | $A$(eV) | $\rho$(Å) | $C$(eV·Å$^6$) | $D$(eV·Å$^{12}$) |
| Si-O | 13702.9050 | 0.193817 | 54.681 | 2.5 |
| Al-O | 12201.4170 | 0.195628 | 31.997 | 1 |
| Na-O | 2755.0323 | 0.258583 | 33.831 | 5 |
| O-O | 2029.2233 | 0.343645 | 192.58 | 150 |
| Lennard-Jones potential parameters | | | |
| Atomic pair | $\varepsilon$(eV) | | $\sigma$(Å) |
| Si-C | 0.004161092 | | 3.6605 |
| Al-C | 0.003441915 | | 3.741 |
| Na-C | 0.00477315 | | 3.0855 |
| O-C | 0.00532142 | | 3.268 |
| C-C | Optimized Tersoff potential | | |

## 2.3. Thermal conductivity calculation

2.3.1. Non-equilibrium molecular dynamics

The thermal conductivity of CNT-GP nanocomposites was calculated by using the non-equilibrium molecular dynamics method (NEMD). The initial structures were first equilibrated for 400 ps under NVT ensemble, and the temperature was maintained at 300 K using the Nosé-Hoover heat bath. Then, the system was further equilibrated under an NVE ensemble. In order to obtain the temperature distribution, the system was divided into N (N=21) blocks along the heat flux direction, and the temperature of the heat source and heat sink were maintained at 350 K and 250 K, respectively, using the Langevin heat bath. The total simulation time was set as 2 ns to make the temperature distribution reach a steady state. The heat flux $J$ can be obtained by calculating the rate ($dE/dt$) of adding or removing energy to the heat sink or source regions, respectively, which is expressed as

$$J = \frac{dE/dt}{A}, \qquad (3)$$

where $A$ is the cross-sectional area in the direction perpendicular to the heat flux. In the statistical stage, we sample the accumulated energy exchange between the system and the thermal baths (as shown in Figure 2a) and the local temperatures (as shown in Figure 2b) for the last 1 ns. The thermal conductivity ($\kappa$) of the system can be obtained by

$$k = \frac{J}{dT/dx}. \qquad (4)$$

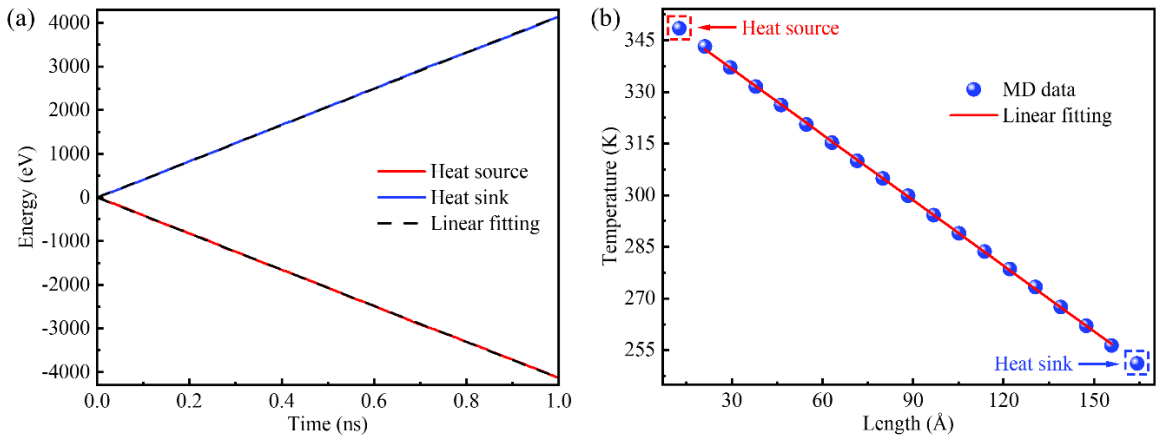

**Figure 2.** (a) The accumulated energy of the Langevin heat bath (heat source and heat sink). (b) Steady-state mean temperature distribution curve for the geopolymer system (GP4)

2.3.2. Spectral thermal conductivity calculation

The spectral heat current (SHC) describes the contribution of phonons with different frequency to heat flux, which is obtained by the calculation of the force and velocity correlation functions near the virtual interface between heat source and sink regions. The spectral decomposition of the heat current $q_{i \to j}(\omega)$ between atoms $i$ and $j$ located at different sides of the virtual interface is calculated as,

$$q_{i \to j}(\omega) = -\frac{2}{t_{simu}\omega} \sum_{\alpha,\beta \in \{x,y,z\}} Im \langle \hat{v}_i^\alpha(\omega)^* K_{ij}^{\alpha\beta} \hat{v}_j^\beta(\omega) \rangle, \tag{5}$$

where $t_{simu}$ is the simulation time, $\omega$ is the angular frequency, $K_{ij}^{\alpha\beta}$ is the force constant matrix which is calculated by,

$$K_{ij}^{\alpha\beta} = \frac{\partial^2 U}{\partial u_i^\alpha \partial u_j^\beta}\bigg|_{u=0}, \tag{6}$$

where $u_i^\alpha$ and $u_j^\beta$ denote the displacements of atoms $i$ and $j$ from their equilibrium positions in the direction of $\alpha, \beta \in \{x, y, z\}$, $U$ denotes the local potential energy. The final $q(\omega)$ through the virtual interface is expressed by [30,31],

$$q(\omega) = \frac{1}{A} \sum_{i \in \tilde{L}} \sum_{j \in \tilde{R}} q_{i \to j}(\omega), \tag{7}$$

where $A$ is the virtual interface area, $\tilde{L}$ and $\tilde{R}$ denote the region within cutoff radius (12 Å) on the left and right sides of the virtual interface, respectively. Based on the SHC ($q(\omega)$), the transmission coefficient ($T(\omega)$) can be calculated by,

$$T(\omega) = \frac{q(\omega)}{k_B \Delta T}, \tag{8}$$

where $k_B$ is the Boltzmann constant, $\Delta T$ denotes the temperature difference between the heat source and heat sink. Based on the SHC ($q(\omega)$), the spectral thermal conductivity (STC) is given by,

$$\kappa(\omega) = \frac{q(\omega)}{A\Delta T}L, \tag{9}$$

where $\Delta T$ and $L$ denote the temperature difference between the heat source and heat sink and the corresponding length, respectively. Thus, the thermal conductivity of the simulated system can also be calculated from the integral of the STC with frequency,

$$\kappa = \int_{-\infty}^{+\infty} k(\omega)d\omega. \tag{10}$$

## 3. RESULTS AND DISCUSSION

### 3.1. Effect of system length on thermal conductivity.

Following the simulation system modeling and thermal conductivity calculation method introduced in Section 2, the thermal conductivity of all systems was evaluated as shown in Figure 3a. The thermal conductivity of GP system without CNT (0% CNT) keeps almost constant when the system length increases, which indicates that the pure GP system has no obvious size effect. With the increase of the system length, the thermal conductivity of CNT-GP nanocomposites along the CNT axial direction always shows obvious increasing trend at 300 K for different number or content of CNT, which is consistent with the trend in other literatures[32,33]. This size effect is mainly due to the size effect of CNT in which the phonon mean free path (MFP) of CNT is around 200 nm[34]. Within the system size range investigated in this work, the phonon transportation in CNT is almost ballistic. With the increase of the system length, the thermal conductivity of CNT increases almost linearly, [30] which in turn leads to obvious thermal conductivity increase in the CNT-GP nanocomposites. Take the 3.3% CNT-GP as example, when the system length increases from 34 Å to 135 Å, the obtained thermal conductivities are 2.82, 3.73, 4.63 and 5.38 W/(m·k), respectively. In addition, the significant size effect of materials at micro-nano scale is also related with the phonon scattering effect at the heat source and heat sink[35,36]. Increasing the system length can reduce the scattering effect, thereby promoting the increase of the final thermal conductivity. In order to reduce the size effect and to predict the thermal conductivity of the CNT-GP system at infinite size, a linear fit can be performed as a function of the inverse of the thermal conductivity ($1/k$) and the inverse of the finite length ($1/L$), [35]

$$\frac{1}{\kappa} = \frac{1}{\kappa_\infty} \times (1 + \frac{\lambda}{L}), \tag{11}$$

where $k_\infty$ represents the thermal conductivity at infinite length, $\lambda$ represents the effective phonon mean free path (MFP). Figure 3b shows the intrinsic thermal conductivity for the 3.3% CNT-GP systems, which is about 9.56 W/(m·k).

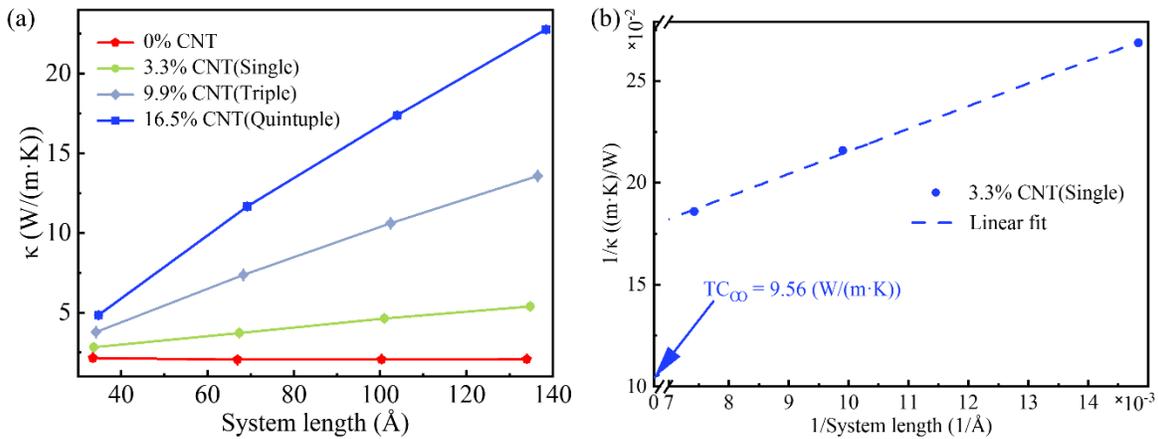

**Figure 3.** (a) Thermal conductivity ($k$) of CNT-GP systems with different length. (b) The variation of $1/k$ with $1/L$ for the 3.3% CNT-GP systems and the thermal conductivity inferred from a linear fit for the infinite length system.

In addition, the increase of thermal conductivity of the multiple CNT-GP systems compared with the single CNT-GP systems is shown in Figure 4. When the system length is longer than 70 Å, the thermal enhancement ratios are larger than the corresponding CNT doping content ratios (3 times and 5 times, respectively). This indicates that there is coupling interaction between two neighboring CNTs, which will be beneficial to the heat transport in CNT-GP nanocomposites.

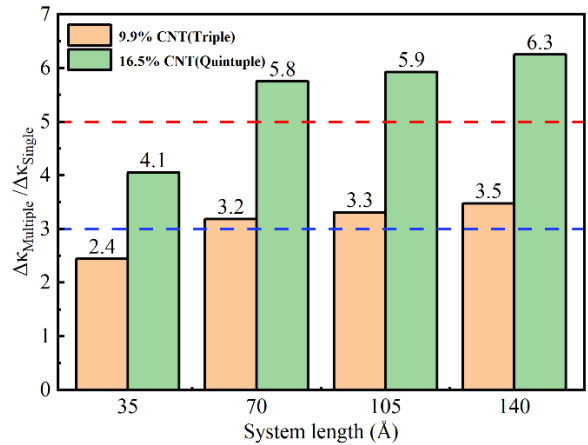

**Figure 4.** The relationship between the increment of thermal conductivity of the multiple CNT systems ($\Delta\kappa_{Multiple}$) divided by the single CNT systems ($\Delta\kappa_{Single}$) and the system length.

In order to explore the intrinsic mechanism of the size-dependent thermal conductivity of CNT-GP nanocomposites, we further calculated the phonon transmission coefficient, spectral thermal conductivity and the corresponding accumulated thermal conductivity, respectively. It can be seen from Figure 5a that the transmission coefficient of the CNT-GP system is more length dependent in the low frequency range than that of the high frequency range. The phonon transportation is mainly ballistic where the system size is smaller than phonon MFP especially in CNT, and the thermal conductivity increases linearly with the increase of the system length. Figure 5b shows the spectral decomposition curve of thermal conductivity for systems with different length. It can be seen that in the low frequency range (<5 THz), the thermal conductivity increases few with the increase of the system length, while in the high frequency range (>5 THz), the thermal conductivity increases significantly with the system length. When the system length increases gradually from 34 Å to 135 Å, the accumulated thermal conductivities of the CNT-GP nanocomposites are 2.55, 3.53, 4.30 and 5.22 W/(m·k), respectively, which is consistent with the NEMD calculation results. In addition, the anharmonic phonon-phonon interactions included in NEMD makes the transmission coefficient dependent on the system length,[37] which can be expressed by the following equation.

$$T(\omega) = \frac{M(\omega)}{1+L/\Lambda(\omega)}, \qquad (12)$$

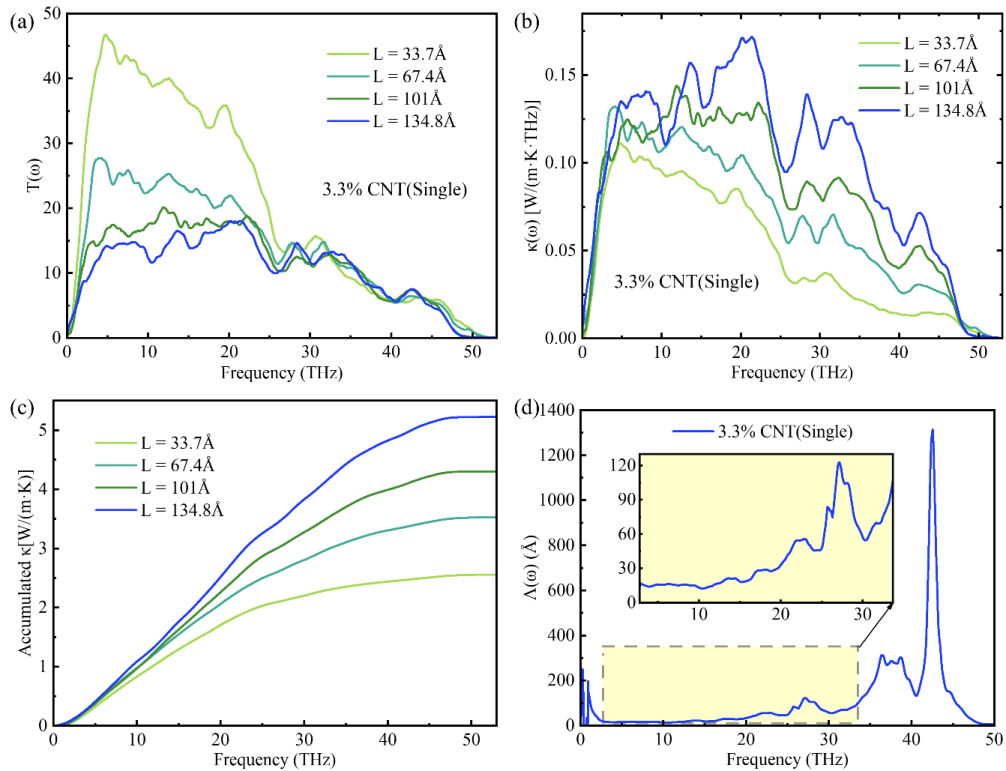

**Figure 5.** (a-c) The transmission coefficient, spectral thermal conductivity and the corresponding accumulated thermal conductivity curves with frequency for 3.3% (Single) CNT systems

(CNT/GP1-4), respectively. (d) The relationship between the effective phonon MFP and frequency for CNT/GP1-4 system.

where $M(\omega)$ is the number of propagating modes at frequency $\omega$.[38] $\Lambda(\omega)$ is the effective phonon MFP at frequency $\omega$. In this study, $M(\omega)$ is approximately equal to the transmission coefficient in the ballistic limit. The effective length of the system used to calculate the ballistic limit transmission coefficient is 0.89 nm. Combining Eqs. 8 and 12, by linearly fitting the data points of $M(\omega)/T(\omega)$ and $L$, the inverse of the slope is $\Lambda(\omega)$. The calculated results in Figure 5d shows that the MFP of the system is larger in the high frequency region than that of the low frequency region, which means the high frequency vibrations contribute most to the total thermal conductivity.

### 3.2. Anisotropic thermal conductivity of CNT-GP nanocomposites

In addition to the system length, the content of CNT also has significant effect on the thermal conductivity. Therefore, four systems with different CNT contents, 0% CNT, 3.3% CNT(Single), 9.9% CNT(Triple) and 16.5% CNT(Quintuple), respectively. In addition, since CNT is anisotropic, its thermal conductivity and thermal transport mechanism in the axial direction and vertical to the axial direction differ greatly, we will study them separately from two aspects below.

#### 3.2.1. Thermal conductivity along CNT axial direction.

It can be seen from the results (Figure 6a) that when the content of CNT is gradually increased from 0 % to 16.5 %, the thermal conductivity of composite systems in the CNT axial direction is 2.15, 2.82, 3.78 and 4.85 W/(m·k), respectively. When the content of CNT is 16.5 %, its thermal conductivity increases by 125.6 % compared with pure GP. To further explore its intrinsic mechanism, we calculated the phonon density of states (Figure 6b), the phonon participation ratio (Figure 6c) and the corresponding STC (Figure 6d) curves for different systems, respectively. The phonon density of states (PDOS) is useful information for characterizing phonon activities in materials.[39] In MD, the PDOS is obtained by calculating the velocity autocorrelation function of atoms in the system and then performing Fourier transform, [40]

$$PDOS(\omega) = \int_{-\infty}^{\infty} e^{i\omega t} \langle \sum_{j=1}^{N} v_j(t) \cdot v_j(0) \rangle dt, \tag{13}$$

where $v_j(t), v_j(0)$ are the velocities of atom $j$ at moments $t$ and 0, respectively; $\omega$ is the vibrational frequency; $N$ is the total number of atoms in the system; PDOS($\omega$) is the PDOS at the vibrational frequency $\omega$. The phonon participation ratio is another effective method to gain insight into phonon activities information, which can be obtained directly from MD simulation without lattice dynamics calculation, greatly improving the computational efficiency. Its expression is

$$PPR(\omega) = \frac{1}{N} \frac{\left(\sum_i PDOS_i(\omega)^2\right)^2}{\sum_i PDOS_i(\omega)^4}, \tag{14}$$

where $N$ is the total number of atoms in the system; $PDOS_i(\omega)$ is the PDOS of atom $i$ at the vibrational frequency $\omega$. From the PDOS curve and the PPR curve, it can be seen that increasing the content of CNT helps to improve the PDOS and the PPR in the high frequency region (35-55 THz). Although pure GP (red curve in Figure 6c) tends to have a rapidly increasing PPR above 35 THz, the PDOS in this region (red curve in Figure 6b) is almost zero. This indicates that there are almost no high frequency phonons above 35 THz in this system, so the contribution of thermal conductivity in this region should also be zero. It can be seen from the SHC calculated result in Figure 6d that with the increase of CNT content, the thermal conductivity tends to increase in almost all frequency ranges. For the low frequency region (0-20 THz), although the PDOS and the PPR both tend to decrease with increasing CNT content, the thermal conductivity in this region still tends to increase. This is duo to the fact that although the overall density of states and participation ratio of the system decrease, the density of states and participation ratio of carbon atoms increase, indicating that carbon nanotubes contribute more to the thermal conductivity of the system. In the high frequency region (above 35 THz), the thermal conductivity of the system without CNT is almost zero, which is consistent with our previous analysis.

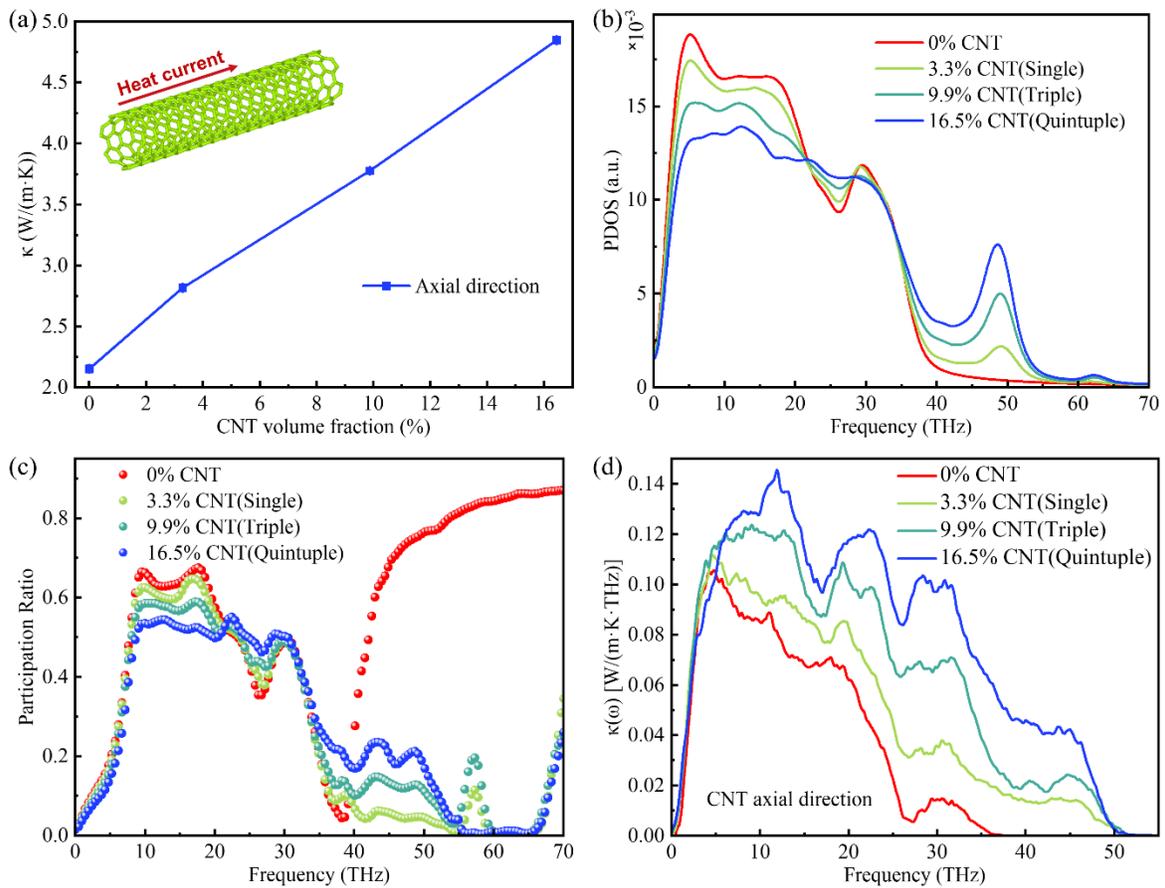

**Figure 6.** (a) The relationship between the thermal conductivity in the CNT axial direction and the content of CNT in systems (GP1, CNT/GP1, CNT/GP5, CNT/GP9). (b-d) The PDOS, PPR and STC curves with frequency for different CNT content systems, respectively.

### 3.2.2. Thermal conductivity perpendicular to CNT axial direction

It can be seen from the results (Figure 7a) that when the content of CNT is gradually increased from 0% to 16.5%, the thermal conductivity of composite systems in the CNT vertical axial direction is 2.15, 1.85, 1.44 and 1.25 W/(m·k), respectively. When the content of CNT is 16.5%, its thermal conductivity vertical to the CNT axial direction is 41.9% lower than the system without CNT. However, there are some CNT composites that also have higher thermal conductivity in the direction of vertical to CNT axial than the system without CNT (such as CNT/epoxy composites[41], etc.). Figure 7b shows the spectral decomposition curve of thermal conductivity in the CNT vertical axial direction. The results show that the thermal conductivity tends to decrease at all frequency ranges as the CNT content increases. In addition, compared with Figure 6d, we can also find that the high frequency phonons above 35 THz only contribute to the heat transport in CNT axial direction, and do not affect the thermal conductivity vertical to the CNT axial direction. To investigate the reason for the decrease in thermal conductivity, we plotted the temperature distribution curves when thermal conductivity was calculated using the NEMD method (Figure 7c–e). The red bolded region is the interface area between carbon nanotubes and geopolymer. It can be found that there is a large temperature gradient in these regions. The equation for calculating the interface thermal resistance is

$$R_{int} = \frac{\Delta T}{q}, \quad (15)$$

where $\Delta T$ is the temperature drop at the interface. $q$ is the heat flux through the interface. Thus, the results show that there is a large interfacial thermal resistance between CNT and GP (the red bolded region), which causes severe phonon scattering of heat carriers in the interface and eventually leads to a reduction in thermal conductivity at all frequency ranges. However, some experimental studies have shown that the use of interface engineering can significantly improve the thermal conductivity of composite materials.[2] For example, the introduction of C-O interactions enable the formation of covalent bonds between CNT and GP, rather than simply van der Waals interactions. This may reduce the interface thermal resistance and phonon scattering at the interface region, which will be further investigated in the future.

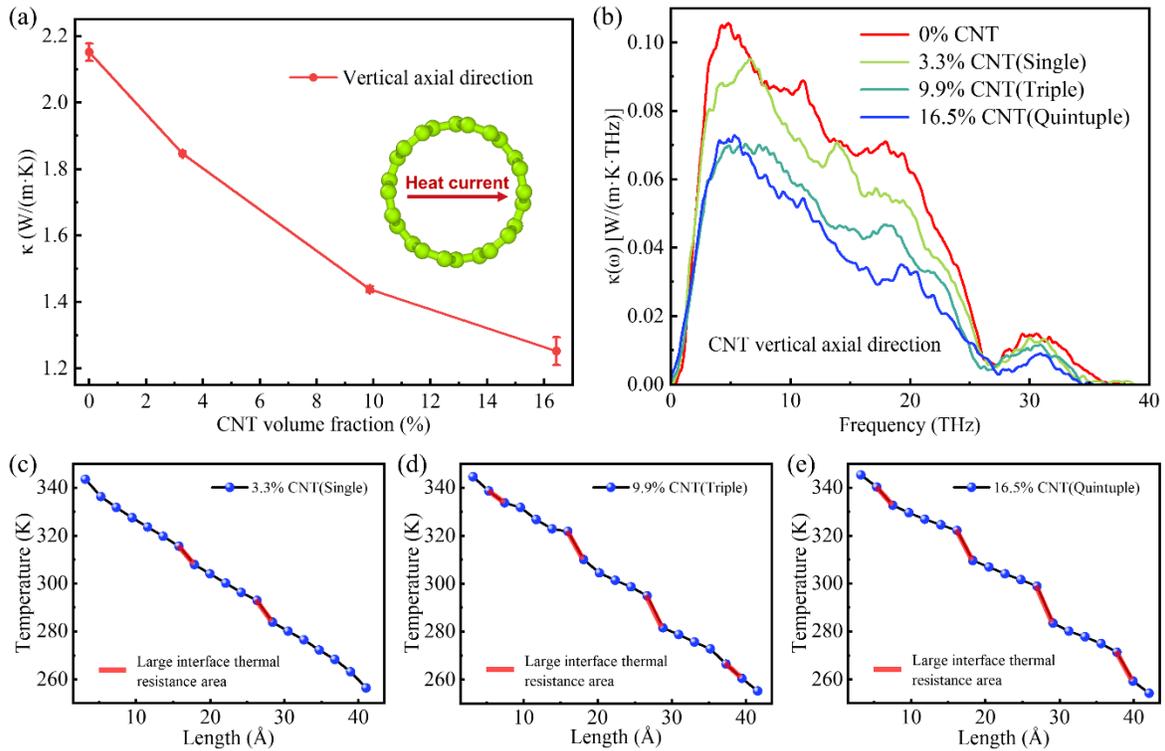

**Figure 7.** (a) The relationship between the thermal conductivity in the CNT vertical axial direction and the content of CNT in systems (GP1, CNT/GP1, CNT/GP5, CNT/GP9). (b) The STC curve with frequency for different CNT content systems. (c-e) Temperature distribution curves for different systems (CNT/GP1, CNT/GP5, CNT/GP9) when thermal conductivity in the CNT vertical axial direction was calculated using the NEMD method.

### 3.3. The effect of 3D CNT networks on thermal conductivity

Considering that carbon nanotubes are actually distributed in the GP with 3D network style, it is necessary to properly expand the simulation system and to study the 3D CNT-GP composite system, which is necessary to guide the experimental research. In the 3D CNT system, the angle between the axial direction of CNT and the direction of heat flux is a key factor affecting the thermal conductivity of the system. We define the mean angle between the axial direction of all carbon nanotubes and the heat flux direction as the mean angle of CNT and heat flux (MACH), which is calculated as follows,

$$MACH = \frac{1}{N}\sum_{i=1}^{N} \theta_i, \tag{16}$$

where $N$ is the total number of CNT in the system. $\theta_i$ is the angle between the axial direction of the $i'th$ CNT and the heat flux direction. Then we constructed some systems (system size: 88 Å×88 Å×88 Å; CNT content: 7.46%; MACH: 0°, 55.9°, 90°, respectively) according to the method in section 2.2. The detailed systems parameters are shown in Table 3. And the length of each carbon nanotube is 55 Å.

Figure 8a shows the structure of the geopolymer composite system with MACH=55.9°. Figure 8b shows the thermal conductivity curve calculated for different systems. The thermal conductivity of the composite systems were 2.78, 1.73 and 1.58 W/(m·k), respectively, when the MACH was gradually increased from 0° to 90°. In addition, the thermal conductivity of the system without CNT can be calculated to be 2.08 W/(m·k). The curve fit was found to satisfy the exponential relationship, and the expression is: $y = 1.28e^{-x/31.8} + 1.5$. When the MACH is in the range of (0°, 25.14°), the addition of carbon nanotubes can promote the thermal conductivity of the composite material, and when the MACH is greater than 25.14°, the addition of carbon nanotubes can inhibit the thermal conductivity of the composite material. The result shows that CNT contribute to the thermal conductivity of the composite material in only 27.93% of the range. This is mainly due to the fact that only van der Waals interactions were considered between CNT and GP in this study, and covalent bonding interactions were not considered. Therefore, there will be large interfacial thermal resistance and phonon scattering, which will significantly affect the heat transport. This can improve the thermal conductivity of composites through the "interface engineering" method described in Section 3.2.2. In addition, due to the limitation of computing resources, the length of CNT in this system is only 5.5 nm, but the thermal conductivity of CNT in the micron scale can reach thousands of W/(m·k). Therefore, the size effect is also an important reason to limit the increase of its thermal conductivity.

**Table 3.** Simulation systems parameters

| Name | System size | CNT content (%) | MACH (°) | Atoms |
|---|---|---|---|---|
| GP5 | 88Å×88Å×88Å | 0 | - | 50648 |
| CNT/GP13 | 88Å×88Å×88Å | 7.46 | 0 | 54407 |
| CNT/GP14 | 88Å×88Å×88Å | 7.46 | 55.9 | 54441 |
| CNT/GP15 | 88Å×88Å×88Å | 7.46 | 90 | 54407 |

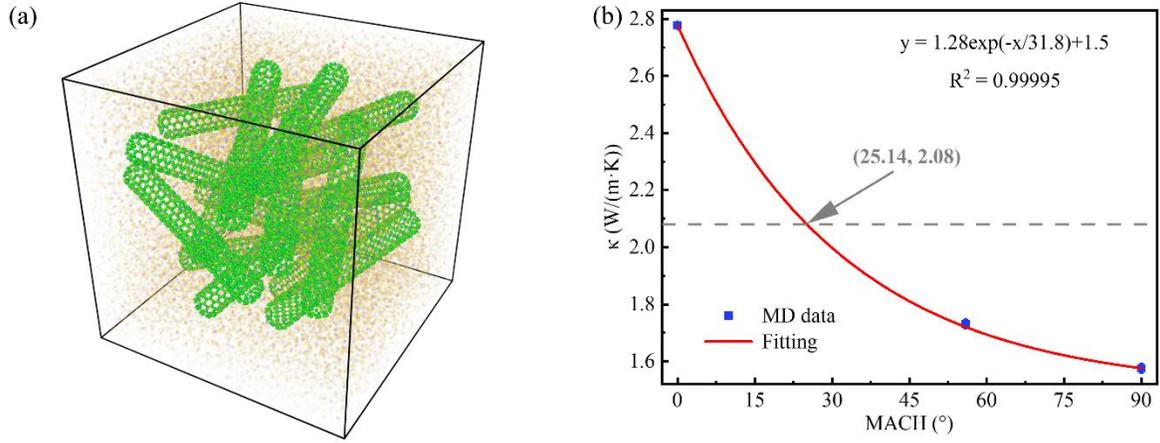

**Figure 8.** (a) Structure diagram of CNT/GP composite system (MACH=55.9°). (b) The relationship between thermal conductivity and MACH for different systems (GP5, CNT/GP13-15).

## 4. CONCLUSIONS

In summary, we have systematically investigated the effect of the size, content and distribution of carbon nanotubes on the thermal conductivity of geopolymer composites using a non-equilibrium molecular dynamics method, and the microscopic thermal conductivity mechanism is analyzed by combining phonon transport analytical methods. It was found that there is a significant size effect in the geopolymer composite system, and the thermal conductivity increases with the length of the system. For the system doped with 3.3% (Single) carbon nanotubes, the thermal conductivity ($k$) of the system increases from 2.82 W/(m·k) to 5.38 W/(m·k) when the system length increases from 34 Å to 135 Å, an increase of 90.8%. In addition to the system length, the content of carbon nanotubes and the heat flux direction also have a great influence on the thermal conductivity of the composite system. when the content of carbon nanotubes is 16.5%, the thermal conductivity in the CNT vertical axial direction (4.85 W/(m·k)) increases 125.6% compared with the system without CNT (2.15 W/(m·k)). However, the thermal conductivity in the CNT vertical axial direction (1.25 W/(m·k)) decreases 41.9%. This is mainly due to the interfacial thermal resistance and phonon scattering at the interface. For the 3D CNT system, we found that when MACH is less than a certain angle (25.14°), carbon nanotubes are beneficial to improve the thermal conductivity of the composite material; beyond this angle, the addition of carbon nanotubes decreases the thermal conductivity of the composite instead. The "interface engineering" is expected to increase the value of the critical angle. This work provides some insights for studying the microscopic heat transport properties of geopolymer system using molecular dynamics method, and has certain guiding significance for the future research on thermal conductivity improvement and multi-scale simulation of geopolymer composite system.

AUTHOR INFORMATION

**Corresponding Author**

**Shenghong Ju** - China-UK Low Carbon College, Shanghai Jiao Tong University, Shanghai, 201306, China; Email: shenghong.ju@sjtu.edu.cn

**Author**

**Wenkai Liu** - China-UK Low Carbon College, Shanghai Jiao Tong University, Shanghai, 201306, China

**Ling Qin** - SPIC Guizhou Jinyuan Co., Ltd., Guiyang, Guizhou, 550081, China.

**C. Y. Zhao** - China-UK Low Carbon College, Shanghai Jiao Tong University, Shanghai, 201306, China

**Notes**

The authors declare no competing financial interest.

ACKNOWLEDGMENT

This work was supported by the National Natural Science Foundation of China (No. 52006134), Shanghai Key Fundamental Research Grant (No. 21JC1403300), and the National Key R&D Program of China (2021YFB3702303). The computations in this paper were run on the π 2.0 cluster supported by the Center for High Performance Computing at Shanghai Jiao Tong University.